\documentclass[sigconf]{acmart}

\usepackage{booktabs} 

\usepackage{balance}

\def\BibTeX{{\rm B\kern-.05em{\sc i\kern-.025em b}\kern-.08emT\kern-.1667em\lower.7ex\hbox{E}\kern-.125emX}}

\usepackage{subcaption}
\usepackage{float}

\copyrightyear{2020}
\acmYear{2020}
\setcopyright{acmlicensed}\acmConference[SIGIR '20]{Proceedings of the 43rd International ACM SIGIR Conference on Research and Development in Information Retrieval}{July 25--30, 2020}{Virtual Event, China}
\acmBooktitle{Proceedings of the 43rd International ACM SIGIR Conference on Research and Development in Information Retrieval (SIGIR '20), July 25--30, 2020, Virtual Event, China}
\acmPrice{15.00}
\acmDOI{10.1145/3397271.3401220}
\acmISBN{978-1-4503-8016-4/20/07}

\author{Casper Hansen}
\authornote{Both authors share the first authorship}
\affiliation{
  \city{University of Copenhagen}
}
\email{c.hansen@di.ku.dk}

\author{Christian Hansen}
\authornotemark[1]
\affiliation{
  \city{University of Copenhagen}
}
\email{chrh@di.ku.dk}

\author{Jakob Grue Simonsen}
\affiliation{
  \city{University of Copenhagen}
}
\email{simonsen@di.ku.dk}

\author{Stephen Alstrup}
\affiliation{
  \city{University of Copenhagen}
}
\email{s.alstrup@di.ku.dk}

\author{Christina Lioma}
\affiliation{
  \city{University of Copenhagen}
}
\email{c.lioma@di.ku.dk}

\begin{document}
\fancyhead{}
\begin{abstract}
Semantic Hashing is a popular family of methods for efficient similarity search in large-scale datasets. In Semantic Hashing, documents are encoded as short binary vectors (i.e., hash codes), such that semantic similarity can be efficiently computed using the Hamming distance. Recent state-of-the-art approaches have utilized weak supervision to train better performing hashing models. Inspired by this, we present Semantic Hashing with Pairwise Reconstruction (PairRec), which is a discrete variational autoencoder based hashing model. PairRec first encodes weakly supervised training pairs (a query document and a semantically similar document) into two hash codes, and then learns to reconstruct the same query document from both of these hash codes (i.e., pairwise reconstruction). This pairwise reconstruction enables our model to encode local neighbourhood structures within the hash code directly through the decoder. We experimentally compare PairRec to traditional and state-of-the-art approaches, and obtain significant performance improvements in the task of document similarity search.
\end{abstract}


\title{Unsupervised Semantic Hashing with Pairwise Reconstruction}
\maketitle

\section{Introduction} \label{s:introduction}
Document similarity search is a core information retrieval task, where semantically similar documents are retrieved based on a query document. Large-scale retrieval requires methods that are both effective and efficient, and that can--ideally--be trained in an unsupervised fashion due to the high cost associated with labeling massive data collections. To this end, Semantic Hashing \cite{salakhutdinov2009semantic} methods learn to transform objects (e.g., text documents) into short binary vector representations, which are called \emph{hash codes}.
The semantic similarity between two documents can then be computed using the Hamming distance, i.e., the sum of differing bits between two hash codes, which can be implemented highly efficiently on hardware due to operating on fixed-length bit strings (real-time retrieval among a billion hash codes \cite{shan2018recurrent}). Hash codes are typically the same length as a machine word (32 or 64 bits), thus the storage cost for large document collections is relatively low. 

The state-of-the-art on unsupervised semantic hashing uses \emph{weak supervision} in different ways to learn hash codes that better encode the structure of local neighbourhoods around each document. NbrReg \cite{chaidaroon2018deep} used BM25 to associate each document with an aggregation of the most similar neighbourhood documents, where two different decoders are trained to reconstruct the document hash code to both the original \textit{and} aggregated neighbourhood document. 
However, we argue that using multiple different decoders on a single hash code is ineffective, since each decoder will attempt to enforce (potentially) different semantics, which may harm generalization of the hash code. Additionally, an aggregated neighbourhood document is not a \textit{real} document encountered during retrieval, which means that learning from it can introduce further semantic shift. Recently, RBSH \cite{hansensemhash2019} proposed to use weak supervision for incorporating a ranking objective in the model, with the aim of improving the hash codes performance in document ranking tasks. However, RBSH uses two weakly (positively and negatively) labeled documents to generate a ranking triplet, each of which is obtained from a noisy relevance estimate, which may lead to larger inaccuracies when combined.


To address the above problems, we propose to use weak supervision to extract the top-K most similar documents to a given query document, which are split into K pairs, each consisting of the query document and a top-K document. Using an end-to-end discrete variational autoencoder architecture, each document within a pair is encoded to a hash code, and through a single decoder they are both trained in an unsupervised fashion to be able to reconstruct the query document (i.e., they are \emph{pairwise} reconstructed to obtain the query document). In contrast to NbrReg \cite{chaidaroon2018deep}, our PairRec aims at learning a more generalizable decoding through a single decoder used on pairs of (non-aggregated) documents, as opposed to using different decoders as done in NbrReg. In contrast to RBSH \cite{hansensemhash2019}, our PairRec is only based on a single weakly labeled document per sample, thus aiming at reducing the inaccuracies originating from comparing noisy relevance estimates for ranking in RBSH.


In summary, we \textbf{contribute} a novel weakly supervised semantic hashing approach named PairRec, based on our concept of pairwise reconstruction for encoding local neighbourhood structures within the hash code. We experimentally evaluate the effectiveness of PairRec against traditional and state-of-the-art semantic hashing approaches, and show that PairRec obtains significant improvements in the task of document similarity search. In fact, PairRec hash codes generally perform similar or better than the state-of-the-art while using 2-4x fewer bits. 

\section{Related Work}
Early work on semantic hashing used techniques adopted from spectral clustering \cite{weiss2009spectral}, encapsulating global similarity structures, and later local similarity structures between neighbours found using k-nearest neighbour \cite{zhang2010self}. Following the popularity of deep learning, VDSH \cite{chaidaroon2017variational} was proposed as a neural model enabling complex encoding of documents, that aimed to learn more descriptive hash codes. Inspired by the benefit of weak supervision in related domains \cite{dehghani2017neural, hansen2019neural}, NbrReg \cite{chaidaroon2018deep} was proposed for incorporating aggregated neighbourhood documents in the hash code decoder, for the purpose of incorporating local similarity structure. However, these methods do not learn the hash code in an end-to-end fashion, since they rely on a post-processing rounding stage. To this end, NASH \cite{nash2018} was proposed as an end-to-end trainable variational autoencoder, where bits were sampled according to a learned sample probability vector from a Bernoulli distribution. As a step towards more expressive document encoding, BMSH \cite{dong-etal-2019-document} utilized a Bernoulli mixture prior generative model, but was only able to outperform a simple version of the NASH model, and not consistently outperform the proposed full version. Lastly, RBSH \cite{hansensemhash2019} was the first semantic hashing approach that utilized a ranking objective in the model (through sampling semantically similar documents \cite{hansen2019copenhagen}), thus enabling the hash codes to combine both local and global structures for improved retrieval performance. RBSH was able to significantly outperform existing state-of-the-art semantic hashing approaches. Recently, semantic hashing has also been successfully applied to the task of cold-start collaborative filtering, where recent advances enabled a better semantic representation of the items \cite{hansen-coldstart-hash-2020}.

\begin{figure}
    \centering
    \includegraphics[width=0.54\linewidth]{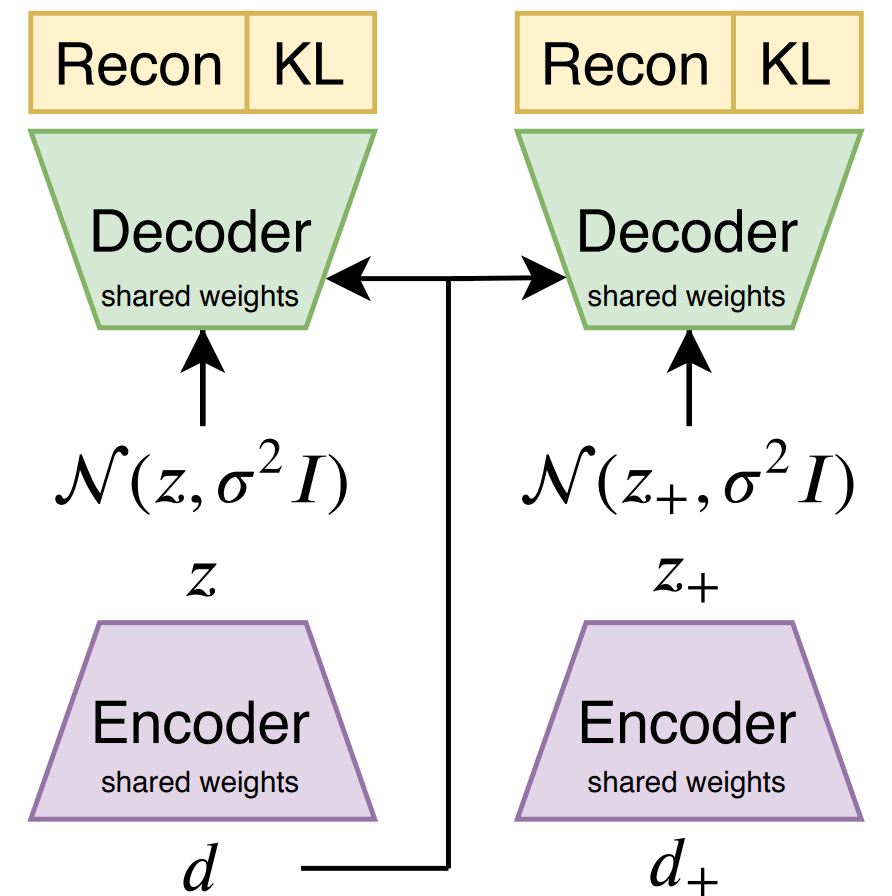}
    \vspace{-10pt}
    \caption{PairRec model overview.
    }
    \label{fig:model}
    \vspace{-15pt}
\end{figure}

\section{Pairwise Reconstruction based Hashing}
Pairwise reconstruction based hashing (PairRec) is a discrete variational autoencoder with a pairwise reconstruction loss.
Given a document $d$, PairRec generates an $m$-bit hash code $z \in \{0,1\}^m$ for $d$, such that two semantically similar documents have low Hamming distance. Specifically, $z$ is sampled by repeating $m$ consecutive Bernoulli trials based on learned sampling probabilities.
Given a similarity function, PairRec is trained on pairs of semantically similar documents, and learns to encode local document neighbourhood structures by training to reconstruct one of the documents from both hash codes (i.e., pairwise reconstruction). We first cover the model architecture and then the pairwise reconstruction loss function. Figure \ref{fig:model} shows a model overview.

To compute the hash code $z$, we let the document likelihood be conditioned on $z$ and define the conditional document likelihood as a product over word probabilities:
\begin{align}\label{eq:doc_likelihood}
   p(d|z) = \prod_{j \in \mathcal{W}_d} p(w_j|z) 
\end{align}
where $\mathcal{W}_d$ denotes the set of all unique words in document $d$. Based on this, the document log likelihood can be found as:
\begin{align} 
    \log p(d) = \log \sum_{z \in \{0,1\}^m} p(d|z)p(z)
\end{align}
where $p(z)$ is the hash code prior of a Bernoulli distribution with equal probability of sampling 0 and 1. However, maximizing $\log p(d)$ is intractable in practice \cite{kingma2013auto}, so instead we maximise the variational lower bound:
\begin{align}\label{trick1}
    \log p(d) \geq E_{Q(\cdot|d)}[\log p(d|z)] - \textrm{KL}( Q(z|d) || p(z))
\end{align}
where $Q(z|d)$ is a learned approximation of the posterior distribution, and KL is the Kullback-Leibler divergence, which has a closed form solution for Bernoulli distributions \cite{nash2018}. Next, we cover our model's encoder ($Q(z|d)$) and decoder ($p(d|z)$), and subsequently specify the pairwise reconstruction loss. 

\subsection{Encoder}\label{s:encoder}
The approximate posterior $Q(z|d)$ is computed using a feedforward network with two hidden layers with ReLU activations, and a final output layer using a sigmoid activation to get the sampling probability for each bit:
\begin{align}
Q(z|d) = \text{FF}_{\sigma}(\text{FF}_{\textrm{ReLU}}(\text{FF}_{\textrm{ReLU}}(d \odot e_{\text{imp}}))))
\end{align}
where $\text{FF}$ denotes a single feed forward layer, $\odot$ is elementwise multiplication, and $e_{\text{imp}}$ is a learned word level importance \cite{hansensemhash2019}. During training, the bits are Bernoulli sampled according to their sampling probabilities, while the most probable bits are chosen greedily for evaluation. This enables exploration during training, and a deterministic evaluation output. As the sampling is non differentiable, the straight through estimator is used to do back propagation through the sampling \cite{bengio2013estimating}.

\subsection{Decoder}
The decoder should reconstruct the original document $d$. Previous work has shown a single linear projection works well \cite{nash2018,hansensemhash2019} because the hash codes are used for (linear) Hamming distance computations. We compute the word probabilities by a softmax, where the logit for a single word is given by:
\begin{align}
    \text{logit}(w|z) = f(z)^T (E_{\textrm{word}} (I(w) \odot e_{\textrm{imp}})) + b_w
\end{align}
where $f(z)$ is a noise infused hash code, $E_{\textrm{word}}$ is a word embedding learned during training,  $I(w)$ is a one-hot encoding of word $w$, $e_{\textrm{imp}}$ is the word level importance also used in the encoder, and $b_w$ is a word level bias term. The noise infusion is done by adding Gaussian noise with zero mean and variance $\sigma^2$ to the hash code, resulting in lower variance for the gradient estimates \cite{kingma2013auto}. We apply variance annealing to reduce the variance over time while training the model. Thus, the conditional document log likelihood is given by:
\begin{align}\label{eq:decoder_function}
    \log p(d|z) = \sum_{j \in \mathcal{W}_d} \log \frac{e^{\text{logit}(w_j|z)}}{e^{\sum_{i \in \mathcal{W}_{\textrm{all}}} \text{logit}(w_i|z)}}
\end{align}
where $\mathcal{W}_{\textrm{all}}$ is the set of unique words over all documents.

\subsection{Pairwise Reconstruction}
PairRec assumes access to some similarity function, which given a document $d$ can be used to obtain a set of the $K$ most similar documents $\mathcal{D}_d^K$. A training pair $(d,d_+)$ is constructed from the document $d$ and a single document sampled from the set, i.e., $d_+ \in \mathcal{D}_d^k$. Using the variational lower bound from Eq. \ref{trick1}, the pairwise reconstruction loss for the pair is defined as:
\begin{align}
    \mathcal{L}_{\textrm{PairRec}} =& -E_{Q(\cdot |d)}[\log p(d|z)] + \beta\textrm{KL}( Q(z|d) || p(z)) \nonumber \\
    -& E_{Q(\cdot |d_+)}[\log p(d|z_+)] + \beta\textrm{KL}( Q(z_+|d_+) || p(z_+))
\end{align}
Note that this is a negation of the variational lower bound because the loss needs to be minimized. The loss consists of two parts: (i) the first part is an ordinary variational lower bound for document $d$; (ii) in the second variational lower bound, document $d_+$ is used in the encoding, while the decoding is of document $d$. This transfers local neighbourhood structure from the document space into the Hamming space, since $z_+$ needs to be able to reconstruct the original $d$. Lastly, the KL divergence is weighed by a tuneable parameter.

\section{Experimental Evaluation}
We use 3 publicly available datasets commonly used in related work \cite{hansensemhash2019, nash2018, chaidaroon2017variational} consisting of TMC, reuters, and agnews (see Table \ref{tab:datasets}). \textit{TMC}\footnote{\url{https://catalog.data.gov/dataset/siam-2007-text-mining-competition-dataset}} is a multi-class dataset of air trafic reports. \textit{reuters}\footnote{\url{http://www.nltk.org/book/ch02.html}} is a multi-class dataset of news, and filtered such that a document is removed if none of its labels are among the 20 most frequent labels (similarly done by \cite{hansensemhash2019, nash2018, chaidaroon2017variational}). Lastly, \textit{agnews} \cite{zhang2015character} is a single-class dataset of news. 

We use the preprocessed data provided in \cite{hansensemhash2019}, where TF-IDF is used as the document representation and words occurring only once are removed, as well as words occurring in more than 90\% of the documents. The datasets were split into training, validation, and testing (80\%/10\%/10\%). We use the validation loss to determine when to stop training a model (using early stopping with a patience of 5 epochs).

\begin{table}[]
    \centering
        \caption{Dataset statistics}
        \vspace{-12pt}
    \resizebox{\linewidth}{!}{
    \begin{tabular}{lcccc}
        \toprule
         & documents & multi-class & classes & unique words \\ \hline
         TMC & 28,596 & Yes &  22 & 18,196 \\
         reuters & 9,848 & Yes & 90 & 16,631 \\
         agnews & 127,598 & No & 4 & 32,154 \\
        \bottomrule
    \end{tabular}
    }
    \label{tab:datasets}
    \vspace{-13pt}
\end{table}
\begin{table*}[]
    \centering
        \caption{Prec@100 with different bit sizes. Bold numbers highlights the highest scores, and $^{\blacktriangle}$ represents statistically significant improvements over RBSH (the best baseline) at the 0.05 level using a two tailed paired t-test.}
        \vspace{-9pt}
    \resizebox{0.9\linewidth}{!}
     {
    \begin{tabular}{l|ccccc|ccccc|ccccc}
    \toprule
    & \multicolumn{5}{c}{Agnews} & \multicolumn{5}{c}{Reuters} & \multicolumn{5}{c}{TMC} \\
    & 8 & 16 & 32 & 64 & 128 & 8 & 16 & 32 & 64 & 128 & 8 & 16 & 32 & 64 & 128 \\ \midrule
SpH \cite{weiss2009spectral} & .3596 & .5127 & .5447 & .5265 & .5566 & .4647 & .5250 & .6311 & .5985 & .5880 & .5976 & .6405 & .6701 & .6791 & .6842\\
STH  \cite{zhang2010self} & .6573 & .7909 & .8243 & .8377 & .8378 & .6981 & .7555 & .8050 & .7984 & .7748 & .6787 & .7218 & .7695 & .7818 & .7797\\
LCH \cite{zhang2010laplacian} & .7353 & .7584 & .7654 & .7800 & .7879 & .5619 & .6235 & .6587 & .6610 & .6586 & .6546 & .7028 & .7498 & .7817 & .7948\\
VDSH \cite{chaidaroon2017variational} & .6418 & .6754 & .6845 & .6802 & .6714 & .6371 & .6686 & .7063 & .7095 & .7129 & .6989 & .7300 & .7416 & .7310 & .7289\\
NbrReg \cite{chaidaroon2018deep} & .4274 & .7213 & .7832 & .7988 & .7976 & .5849 & .6794 & .6290 & .7273 & .7326 & .7000 & .7012 & .6747 & .7088 & .7862\\
NASH  \cite{nash2018} & .7207 & .7839 & .8049 & .8089 & .8142 & .6202 & .7068 & .7644 & .7798 & .8041 & .6846 & .7323 & .7652 & .7935 & .8078\\ 
RBSH  \cite{hansensemhash2019} & .8066 & .8288 & .8363 & .8393 & .8381 & .7409 & .7740 & .8149 & .8120 & .8088 & .7620 & .7959 & .8138 & .8224 & .8193\\ \hline
PairRec (ours) & \textbf{.8119}$^{\blacktriangle}$ & \textbf{.8354}$^{\blacktriangle}$ & \textbf{.8452}$^{\blacktriangle}$ & \textbf{.8492}$^{\blacktriangle}$ & \textbf{.8498}$^{\blacktriangle}$ & \textbf{.7502}$^{\blacktriangle}$ & \textbf{.8028}$^{\blacktriangle}$ & \textbf{.8268}$^{\blacktriangle}$ & \textbf{.8329}$^{\blacktriangle}$ & \textbf{.8468}$^{\blacktriangle}$ & \textbf{.7656}$^{\blacktriangle}$ & \textbf{.7991}$^{\blacktriangle}$ & \textbf{.8239}$^{\blacktriangle}$ & \textbf{.8280}$^{\blacktriangle}$ & \textbf{.8303}$^{\blacktriangle}$ \\ 
\bottomrule
    \end{tabular}}

    \label{tab:results}
    \vspace{-9pt}
\end{table*}

\subsection{Baselines and Tuning}
We compare our PairRec against traditional post-processing rounding approaches (SpH \cite{weiss2009spectral}, STH  \cite{zhang2010self}, and LCH \cite{zhang2010laplacian}), neural post-processing rounding approaches (VDSH \cite{chaidaroon2017variational} and NbrReg \cite{chaidaroon2018deep}), and neural end-to-end approaches (NASH \cite{nash2018} and RBSH \cite{hansensemhash2019}). NbrReg and RBSH both make use of weak supervision as discussed in Section \ref{s:introduction}. The baselines are tuned as described in their original papers.

In PairRec\footnote{We make our code available at \url{https://github.com/casperhansen/PairRec}.}, we tune the number of hidden units in each encoder layers across $\{500, 1000\}$, and the number of top K reconstruction pairs across $\{$1, 2, 5, 10, 25, 50, 100, 150, 200$\}$. For obtaining the reconstruction pairs, we generate 64 bit STH \cite{zhang2010self} hash codes and retrieve the top K most semantically similar documents (STH was also used by RBSH \cite{hansensemhash2019}). For the KL divergence, we tune $\beta$ from $\{0, 0.01, 0.1\}$. Note that when $\beta=0$ is chosen, it corresponds to removing the regularizing KL divergence from the loss. For the variance annealing, we use an initial value of 1 and reduce it by $10^{-6}$ every iteration (as done in \cite{hansensemhash2019}). Lastly, we use the Adam optimizer with a learning rate of 0.0005.

\subsection{Evaluation Setup}
Following related work \cite{hansensemhash2019,nash2018,chaidaroon2017variational,chaidaroon2018deep}, we evaluate the semantic hashing approaches based on their top 100 retrieval performance using Prec@100 based on the Hamming distance. Given a query document, we define a retrieved document to be relevant if it shares at least one label with the query document (to ensure that we can accommodate the multi-class datasets, where each document may have one or more associated labels).

\subsection{Results} \label{s:results} 
We generate hash codes of $\{8, 16, 32, 64, 128\}$ bits and report Prec@100 in Table \ref{tab:results}. The best performing method for each dataset and bit size is highlighted in bold, and statistically significant improvements (0.05 level) using a two tailed paired t-test are indicated by $^{\blacktriangle}$.

Our PairRec method consistently outperforms all the traditional and state-of-the-art approaches across all datasets on all bit sizes. RBSH, which also utilizes weak supervision for generating ranking triplets, consistently obtains the second best scores, indicating the benefit of weak supervision for semantic hashing. While NbrReg also makes use of weak supervision (for creating aggregated neighbourhood documents), it performs worse than both NASH and RBSH, but generally better than VDSH, to which its architecture is most similar to. 
The absolute Prec@100 increases depend on dataset and bit size, but overall PairRec improves state-of-the-art by 1-4\%, which correspondingly enables PairRec hash codes to generally perform better or similar to state-of-the-art hash codes with 2-4x more bits.

\subsection{Impact of Pairwise Reconstruction} \label{ss:impactPairsRecon}
The primary novelty of PairRec is the introduction of pairwise reconstruction. We study the impact of (i) the performance gain obtained by the pairwise reconstruction, and (ii) the performance variance across a varying number of document pairs.

\begin{figure}
    \centering
    \includegraphics[width=0.8\linewidth]{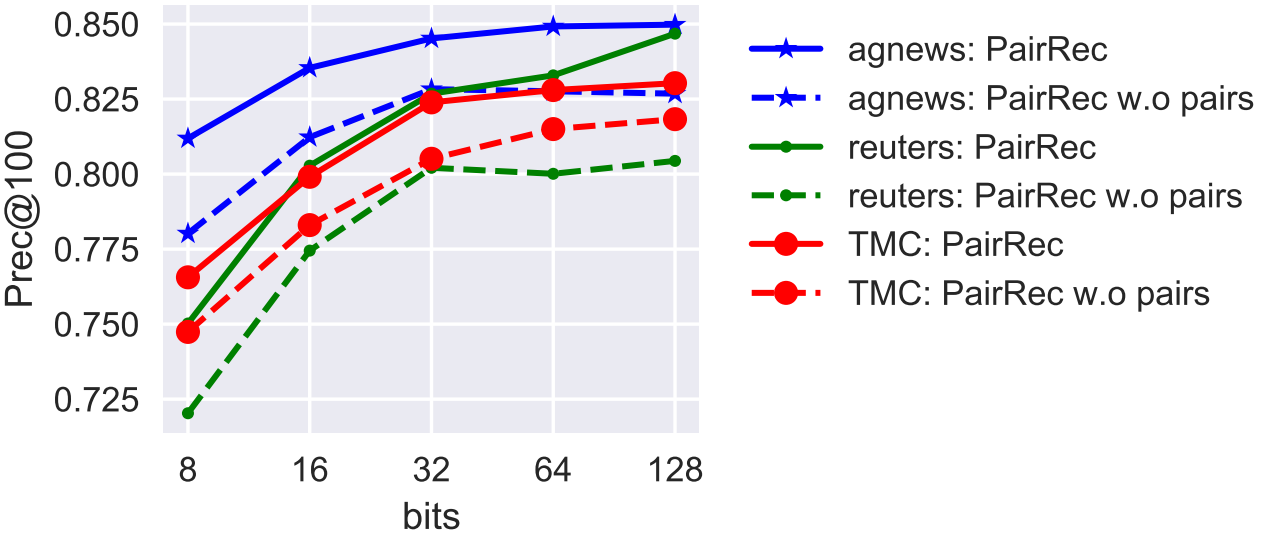}
    \vspace{-13pt}
    \caption{PairRec with and without pairwise reconstruction.}
    \label{fig:pairscomp}

    \vspace{1pt}
    \centering
    \includegraphics[width=0.8\linewidth]{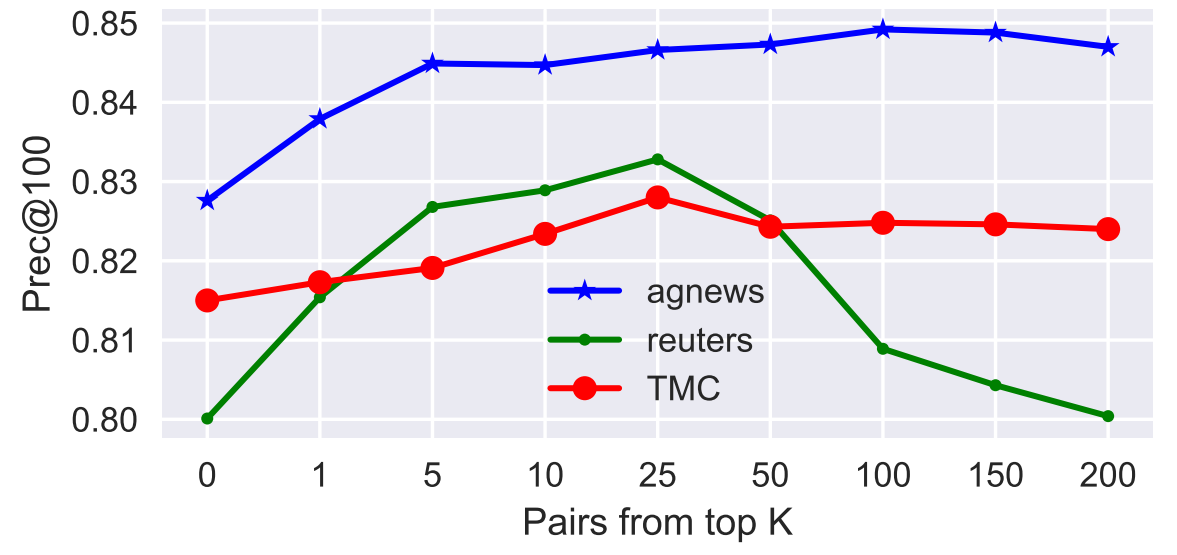}
    \vspace{-13pt}
    \caption{64 bit PairRec while varying the top K.
    }
    \label{fig:nbrcomp}
    \vspace{-15pt}
\end{figure}

\textbf{Performance gain by pairwise reconstruction}.
We compute the Prec@100 with and without the pairwise reconstruction and plot the scores in Figure \ref{fig:pairscomp}. The largest improvements occur for 64-128 bit on the reuters dataset, but across all datasets and bit sizes, pairwise reconstruction obtains consistent improvements. In comparison, the original RBSH paper \cite{hansensemhash2019} also did an ablation with and without weak supervision, but found their improvements to be primarily isolated to 8-16 bits. This further highlights the benefit of using a single weakly supervised document, rather than combining multiple sources for generating ranking triplets as done in RBSH.

\textbf{Performance variance across number of pairs}.
We now investigate the impact of the choice of the number of pairs. We fix the bit size to 64 and plot the Prec@100 for all datasets using $\{$0, 1, 5, 10, 25, 50, 100, 150, 200$\}$ pairs, where 0 corresponds to no pairwise reconstruction. The optimal values for agnews, reuters, and TMC are 100, 25, and 25, respectively. Interestingly, Prec@100 drops after 25 pairs on reuters, which most likely is due to a combination of its small dataset size and high number of classes, corresponding to pairs from top 50 and above no longer being sufficiently semantically similar to the original document. In contrast, for TMC and agnews, we observe no significant performance drop as the number of pairs is increased. In all cases, we note that the optimal value of pairs is also identified by the model parameter configuration with the minimum loss.

\section{Conclusion}
Inspired by recent advances in semantic hashing using weak supervision, we presented a novel semantic hashing approach with pairwise reconstruction (PairRec). PairRec is a discrete variational autoencoder trained on semantically similar document pairs (obtained through weak supervision), where the model is trained such that the hash codes from both pairwise documents reconstruct the same document.
We denote this type of reconstruction as \emph{pairwise reconstruction}; it enables PairRec to encode local neighbourhood structures within the hash code. In an experimental comparison, PairRec was shown to consistently outperform existing state-of-the-art semantic hashing approaches. These improvements generally enable PairRec hash codes to use 2-4x fewer bits than state-of-the-art hash codes while achieving the same or better retrieval performance.

\bibliographystyle{ACM-Reference-Format}
\bibliography{bibfile.bib}
\end{document}